\documentclass[twocolumn]{aastex631}

\usepackage{longtable}

\received{June 5, 2025}
\revised{July 23, 2025}
\accepted{July 30, 2025}

\begin{document}

\title{The Impact of Foregrounds on Dark Ages Measurements with the Highly Redshifted 21\,cm Line}

\correspondingauthor{Jonathan C. Pober}
\email{Jonathan\_Pober@brown.edu}

\author[0000-0002-3492-0433]{Jonathan C. Pober}
\affiliation{Department of Physics, Brown University, Providence, RI 02912-1843}

\author{Willow Smith}
\affiliation{Department of Physics, Brown University, Providence, RI 02912-1843}



\begin{abstract}

Studies of the cosmic dark ages ($30 \lesssim z \lesssim 150$) using the highly redshifted 21\,cm line of neutral hydrogen offer unparalleled amounts of cosmological information, and recent years have seen the refinement of concepts for such experiments (e.g. CoDEX and FarView), nominally feasible with technology and resources in the next one to two decades.  This work studies how the ``foreground wedge'' --- a term in the 21\,cm cosmology literature referring to the contamination of power spectrum modes through the combination of smooth-spectrum foreground emission and the frequency-dependent point spread function of a radio interferometer --- manifests at these very high redshifts.  We find the effect is more significant than at Epoch of Reionization redshifts targeted by current ground-based experiments, with foreground avoidance techniques (which discard all $k$ modes falling within the wedge) typically losing an order of magnitude of sensitivity.  Given the extreme faintness of the 21\,cm signal from the cosmic dark ages and the very high sky temperatures (the dominant source of noise) at low radio frequencies, we conclude that some level of foreground subtraction will be necessary to enable dark ages 21\,cm cosmology with experiments of the scale believed to be achievable in the near term.

\end{abstract}



\section{Introduction} 
\label{sec:intro}

21\,cm cosmology offers the only potential probe of the cosmic ``Dark Ages,'' the period of our Universe's history between the release of the Cosmic Microwave Background (CMB) at redshift $z\approx1100$ and the formation of the first stars at $z\approx30$.  During the dark ages, cosmological hydrogen gas is expected to absorb 21\,cm photons out of the CMB spectrum, leading to an observable signal that can trace the distribution of matter in the Universe down to very small scales.  However, the signal is extremely faint and dwarfed by foreground emission that can be up to 9 orders of magnitude stronger \citep{pritchard_and_loeb_2008}.  Further complicating a detection of the signal is the Earth's ionosphere, which is opaque (or at least highly disruptive) at frequencies $\lesssim 30$\,MHz corresponding to the redshifted 21\,cm line from the dark ages.

Because of the immense challenges of dark ages 21\,cm cosmology, most of the progress developing the technique has been driven by ground-based experiments targeting the Epoch of Reionization (EoR) at $14\gtrsim z\gtrsim6$.  Observations at the corresponding frequencies of 100 to 200\,MHz are still extremely challenging; the foregrounds are still five orders of magnitude stronger than the signal of interest and the ionosphere is still turbulent and refractive (but not reflective).  But concerted efforts by experiments like the Hydrogen Epoch of Reionization Array (HERA; \citealt{deboer_et_al_2017}), the LOw Frequency ARray (LOFAR; \citealt{van_haarlem_et_al_2013}) and the Murchison Widefield Array (MWA; \citealt{tingay_et_al_2013a,wayth_et_al_2018}) have led to upper limits on the 21\,cm signal strength and a number of high-precision analysis techniques used to mitigate foregrounds and other systematics.  In recent years, a number of concepts for a lunar far side radio array for 21\,cm dark ages cosmology have been put forward that build on lessons from ground-based experiments, including including CoDEX \citep{koopmans_et_al_2021}, FarView \citep{polidan_et_al_2024}, and DEX \citep{brinkerink_et_al_2025}.

One of the key paradigms introduced by the 21\,cm EoR community is that of the ``wedge'' and the ``EoR window,'' which we discuss in detail in \S\ref{sec:background}, and the corresponding idea of ``foreground avoidance'' (as opposed to ``foreground subtraction'').  \cite{pober_2015} showed how the foreground avoidance approach (where only modes in the EoR window are used for cosmology) becomes more powerful at redshifts $\sim1-2$ in the post-reionization epoch, as the wedge shrinks and the window becomes bigger.  In this work, we analyze the case of high-redshift cosmic dark ages observations to understand the viability of foreground avoidance for these experiments.

The remainder of the paper is structured as follows.  In \S\ref{sec:background}, we recap the wedge and window paradigm in detail.  In \S\ref{sec:wedge-slope} we discuss the reason the wedge changes with redshift and demonstrate the trends towards higher redshifts.  In \S\ref{sec:sensitivities}, we use the fiducial dark ages 21\,cm array concept from \citet{smith_and_pober_2025} to quantify the impact of the wedge on dark ages observations.  We discuss the principal takeaways of this study in \S\ref{sec:discussion} and conclude in \S\ref{sec:conclusions}.  In all calculations, we use the Planck 2018 cosmological parameters: $\Omega_m = 0.3111, \Omega_\Lambda = 0.6889$, and $h = 0.6766$ \citep{planck_2018_VI}.

\section{Background}
\label{sec:background}

\begin{figure}[h]
    \centering
    \includegraphics[width=\linewidth]{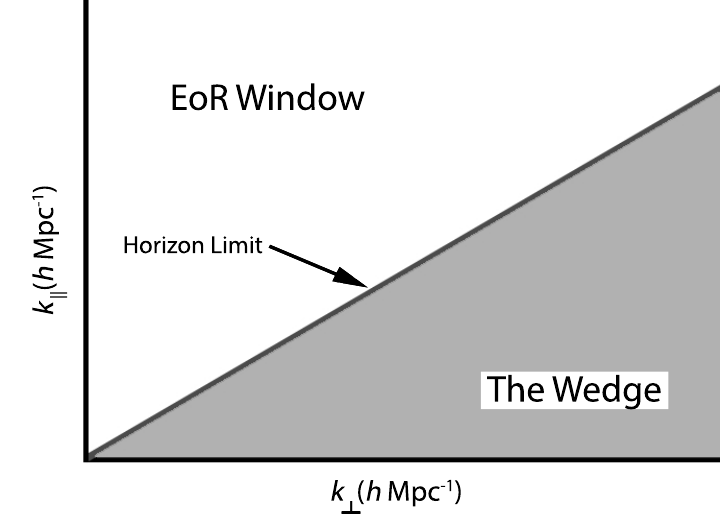}
    \caption{A schematic diagram of the wedge and EoR window in 2D ($k_\perp,k_\parallel$) space.}
    \label{fig:wedge}
\end{figure}

The concept of the foreground wedge and EoR window first appeared in \citet{datta_et_al_2010}, which showed a triangular-shaped region of foreground contamination in simulated 2D ($k_\perp,k_\parallel$) cosmological power spectra of the 21\,cm line; this ``wedge'' is schematically shown in Figure \ref{fig:wedge}.
A number of subsequent papers appeared providing an explanation for this effect (e.g. \citealt{vedantham_et_al_2012,morales_et_al_2012,parsons_et_al_2012b,trott_et_al_2012,thyagarajan_et_al_2013,liu_et_al_2014a,liu_et_al_2014b}) as well as confirmations of the wedge signature in data from early 21\,cm cosmology experiments \citep{pober_et_al_2013b,dillon_et_al_2014,thyagarajan_et_al_2015b}.  These papers were not always in agreement about the cause of the wedge, nor the details of its shape. 
Subsequent work showed that, while the fundamental aspects of the wedge are generic, factors like the array layout \citep{murray_and_trott_2018} and choice of power spectrum estimator \citep{morales_et_al_2019} result in phenomenologically distinct wedge patterns.  In this work, we aim to be agnostic about these details and focus on general aspects of the wedge --- specifically, its slope in ($k_\perp,k_\parallel$) space.  We will return to the impact of the specific estimators used in \S\ref{sec:conclusions}.

\subsection{The Cause of the Wedge}

At the most basic level, the wedge is caused by the frequency-dependent point-spread function (PSF) of a radio interferometer interacting with spectrally smooth foreground emission.  Before we can elaborate on this statement, we will first review how measurements of the 21\,cm line are mapped to the cosmological Fourier space (i.e. $k$-space) used by power spectra.  Following that, we can consider how any foreground contamination in the 21\,cm data ``comes along for the ride'' to cosmological $k$-space and results in the wedge.

21\,cm experiments make measurements vs. frequency (i.e. spectra) with each frequency corresponding to a different redshift of the rest-frame 21\,cm line.  Natively, a radio interferometer measures in $(u,v,f)$-space, where $f$ is the frequency axis, and $u$ and $v$ are the standard $(u,v)$ coordinates of interferometry corresponding the separation vector between pairs of antennas (``baselines'') measured in wavelengths of the observing frequency.  Measurements at different locations in the $(u,v)$ plane approximately correspond to different 2D Fourier modes of the sky brightness distribution.  Cosmological Fourier space ($k_x, k_y, k_z$) is the 3D Fourier transform of a cosmological volume ($r_x,r_y,r_z$), with $x$ and $y$ indicating directions in the plane of the sky and $z$ the direction along the line of sight.  The $(u,v)$ and $(k_x,k_y)$ planes therefore describe the same space and are related by a cosmological scaling that converts (inverse) angles on the sky to (inverse) distances.  With no loss of generality, we can write $\mathbf{k}_\perp \equiv (k_x,k_y)$ and $\mathbf{u} \equiv (u,v)$ and express this relation as \citep{morales_and_hewitt_2004}:
\begin{equation}
\label{eq:kperp}
    \mathbf{k}_\perp = \frac{2\pi\mathbf{u}}{D_C(z)},
\end{equation}
where $D_C(z) \equiv \frac{c}{H_0}\int_0^z\frac{dz'}{E(z')}$ is the line-of-sight comoving distance, $c$ is the speed of light, $H_0$ is the Hubble parameter, and $E(z) \equiv \sqrt{\Omega_m(1+z)^3 + \Omega_\Lambda}$ \citep{hogg99}.  We return to this point in \S\ref{sec:wedge-slope}, but emphasize here that this scaling is \emph{explicitly} redshift-dependent (reflected by the redshift-dependence of the comoving line-of-sight distance): a physical length across the plane of the sky at one redshift subtends a different angle than that same physical length at another redshift.

We can similarly relate the frequency axis $f$ of interferometric measurements to $k_\parallel$, although in a slightly more complicated way.  When observing redshifted 21\,cm line radiation, the observed frequency tells you the redshift, which can be related (through cosmology) to a distance along the line of sight $r_z$.  Therefore, the Fourier transform of $f$ (traditionally called $\eta$ in the literature) can be related to $k_z \equiv k_\parallel$ by \citep{morales_and_hewitt_2004}:
\begin{equation}
\label{eq:kpar}
    k_\parallel = \frac{2\pi\nu_{21}H_0E(z)}{c(1+z)^2}\eta,
\end{equation}
where $\nu_{21}$ is the rest-frame frequency of the 21\,cm line.  We again note that this relationship is explicitly redshift-dependent.

This discussion provides a relatively simple road map for processing a 21\,cm signal measured $(u,v,f)$ space to cosmological $(\mathbf{k_\perp},k_\parallel)$ space: apply a 1D Fourier transform in the $f$ direction, then use equations \ref{eq:kperp} and \ref{eq:kpar} to rescale the coordinates.  The foreground wedge, however, shows up because this procedure \emph{only} works for 21\,cm signals.  A redshifted 21\,cm signal at 100\,MHz (for example) comes from a hydrogen atom located at $z = \nu_{21}/100\,\mathrm{MHz} - 1 = 13.2$.  Foreground emission appears at 100\,MHz because the emission mechanism creating the signal produces electromagnetic radiation at 100\,MHz --- not because of any cosmological redshift.  Equations \ref{eq:kperp} and \ref{eq:kpar} don't make sense when applied to foreground emission, but they are used to rescale the 21\,cm signal in the data, hence our earlier phrase that the foregrounds ``come along for the ride'' when creating a cosmological power spectrum.

However, we are still not quite at the point where we could predict a wedge shape for the foreground contamination in a cosmological power spectrum.  Foreground emission is spectrally smooth, meaning when the Fourier transform from $f$ to $\eta$ is applied, the foregrounds should map primarily to low $\eta$ modes, with no dependence on $u$.  This brings us back to our original statement: the wedge is caused by the frequency-dependent point-spread function (PSF) of a radio interferometer interacting with spectrally smooth foreground emission.  Because the PSF of an interferometer (typically called the ``synthesized beam" in interferometry literature) changes with frequency, signals on the sky acquire additional spectral structure beyond their intrinsic spectral behavior.  By analogy to optical telescopes, the PSF is sometimes said to be ``chromatic'' or that the inteferometer has ``chromaticity".  Moreover, this behavior of the PSF introducing spectral structure has sometimes been called ``mode-mixing'' (because structure in the plane of the sky leads to structure in the frequency direction).  

The final piece of the puzzle for predicting the wedge shape is to note that the small angular scale behavior of the PSF changes more rapidly with frequency than the larger angular scales.  Long baselines (i.e. antenna pairs separated by large distances) of the array probe small angular scales, while short baselines probe large angular scales.  As a function of frequency, the interference pattern of a long baseline changes more rapidly --- long baselines are more chromatic than short baselines.  These are the two facts that give rise to the wedge: long baselines measure large values of $u$ (and therefore large values of $k_\perp$), and long baselines have a more spectrally-structured response.  Hence, the signals measured by long baselines require more $\eta$ modes (and therefore more $k_\parallel$ modes) to describe.  The combination of these two facts is that spectrally smooth signals (like foregrounds), ordinarily expected to occupy only low order $k_\parallel$-modes, bleed upwards in $k_\parallel$, with more bleed at larger values of $k_\perp$.  Several other pedagogical approaches exist for describing the wedge \citep[see e.g.][]{murray_and_trott_2018,liu_and_shaw_2020}, but the basic phenomenology is the same.

\subsection{The Slope of the Wedge}

\begin{figure*}[ht!]
    \centering
    \includegraphics[height=2.75in]{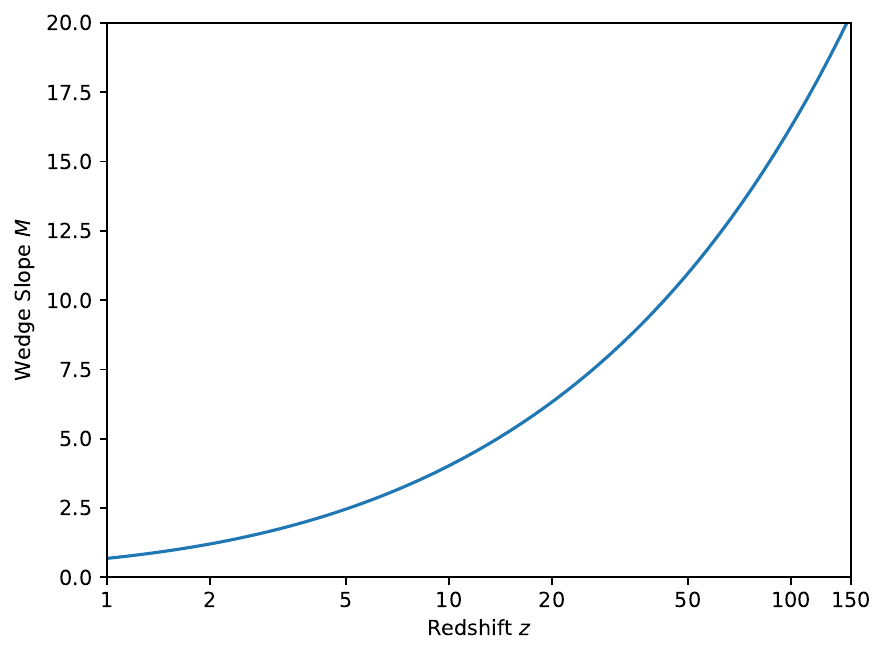}\includegraphics[height=2.75in]{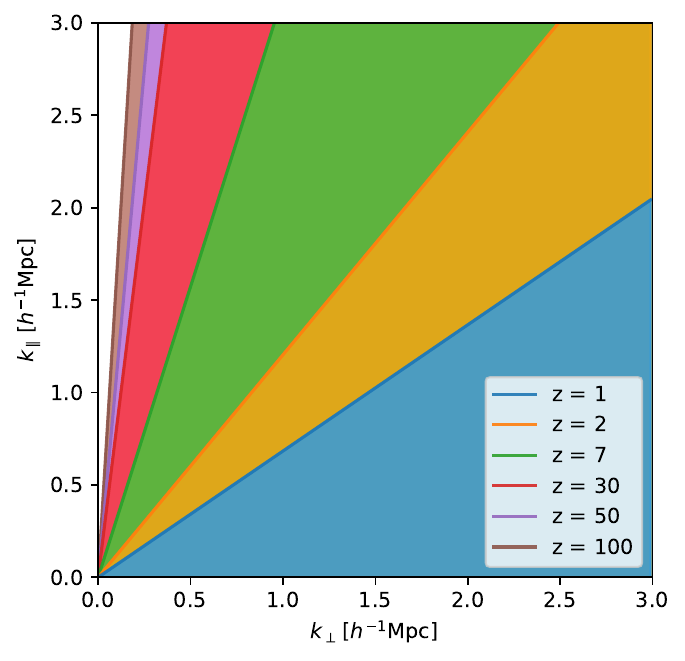}
    \caption{\textit{Left}: The wedge slope $M$ as a function of redshift.  \textit{Right}: The footprint of wedge in 2D $(k_\perp,k_\parallel)$ space for a select number of redshifts. Note that the lower redshift wedges are plotted on top of the higher redshifts, e.g., the wedge at $z=2$ covers both the yellow and blue areas.}
    \label{fig:wedge-slope}
\end{figure*}

The amount of spectral structure in the PSF of an interferometer can be predicted from first principles.  The length of a baseline $\mathbf{b}$, corresponding to a light travel time $\mathbf{b}/c$, sets the maximum rate at which the PSF evolves with frequency at that angular scale.  This maximal amount of spectral structure in turn sets the ``edge'' of the wedge.  \citet{parsons_et_al_2012b} recognized that plane waves from sources on the horizon experience the maximal time delay between their arrival at the two antennas of a baseline, and thus pick up the maximal amount of spectral structure, leading this maximal value to be known as the ``horizon limit'' as indicated in Figure \ref{fig:wedge}.

Analytically, the relationship between baseline length, light-travel time, and the cosmological scalings in equations \ref{eq:kperp} and \ref{eq:kpar} can be used to produce an exact equation for the slope of the wedge \citep{liu_and_shaw_2020}:
\begin{equation}
\label{eq:horizon}
    k_{\parallel,\,\mathrm{horizon}} = k_\perp\frac{H_0D_cE(z)}{c(1+z)},
\end{equation}
where the subscript ``horizon'' indicates this equation applies to the horizon limit, i.e., it defines the upper edge of the wedge.
Again, this equation has an explicit dependence on redshift through $D_C$ and $E(z)$ (as well as the $1+z$ factor in the denominator).  This is the fundamental point underlying the results of this paper: the slope of the wedge changes with redshift.  It was this fact that also drove the results of \citet{pober_2015}: the wedge slope decreased towards low redshifts, increasing the size of the EoR window therefore increasing the number of modes that could be recovered with a foreground avoidance approach (which, again, is defined as only deriving cosmological constraints from modes falling outside the wedge).


\section{Modeling the Redshift Evolution of the Wedge}
\label{sec:wedge-slope}

Using Equation \ref{eq:horizon}, we can straightforwardly study the maximum extent of the wedge as a function of redshift.  Specifically, we study the slope of the horizon line $M$, given by: 
\begin{equation}
\label{eq:wedge-slope}
M \equiv k_{\parallel,\,\mathrm{horizon}}/k_\perp = \frac{H_0D_cE(z)}{c(1+z)}. 
\end{equation}
The left-hand side of Figure \ref{fig:wedge-slope} plots $M$ as a function of redshift; the right-hand side shows the footprint of the wedge in 2D $(k_\perp,k_\parallel)$ space for a select number of redshifts.
It is evident that the wedge slope grows steeper as one approaches higher redshifts (and the corresponding window shrinks).  In \S\ref{sec:sensitivities} we will quantify the loss of modes due to the wedge in terms of the overall detection significance for a fiducial experiment targeting the 21\,cm signal.  We again emphasize that this redshift-dependence in the wedge slope comes from the redshift-dependence of the Hubble parameter and line-of-sight comoving distance, which enter into the scalings necessary to convert angles on the sky and bandwidths along the line of sight to comoving distances.

\section{Sensitivity Predictions}
\label{sec:sensitivities}

\begin{table*}[ht!]
    \centering
    \begin{longtable}{c|cc|cc|cc}
         & \multicolumn{2}{c|}{Fiducial} & \multicolumn{2}{c|}{$10\times$ Sensitivity} & \multicolumn{2}{c}{$100\times$ Sensitivity} \\ 
         Redshift & No FG & FG Wedge & No FG & FG Wedge & No FG & FG Wedge \\
         \hline
        $z = 30$ & $10\sigma$ & $1.1\sigma$ & $100\sigma$ & $11\sigma$ & $1000\sigma$ & $110\sigma$ \\
        $z = 40$ & $6.1\sigma$ & $<1\sigma$ & $61\sigma$ & $6.0\sigma$ & $610\sigma$ & $60\sigma$ \\
        $z = 50$ & $2.5\sigma$ & $<1\sigma$ & $25\sigma$ & $2.4\sigma$ & $250\sigma$ & $24\sigma$ \\
        $z = 60$ & $<1\sigma$ & $<1\sigma$ & $8.1\sigma$ & $<1\sigma$ & $81\sigma$ & $7.7\sigma$ \\
        $z = 70$ & $<1\sigma$ & $<1\sigma$ & $2.6\sigma$ & $<1\sigma$ & $26\sigma$ & $2.3\sigma$ \\   
        $z = 80$ & $<1\sigma$ & $<1\sigma$ & $<1\sigma$ & $<1\sigma$ & $7.7\sigma$ & $<1\sigma$ \\   
        $z = 90$ & $<1\sigma$ & $<1\sigma$ & $<1\sigma$ & $<1\sigma$ & $2.5\sigma$ & $<1\sigma$ \\   
        $z = 100$ & $<1\sigma$ & $<1\sigma$ & $<1\sigma$ & $<1\sigma$ & $<1\sigma$ & $<1\sigma$ \\   
        $z = 110$ & $<1\sigma$ & $<1\sigma$ & $<1\sigma$ & $<1\sigma$ & $<1\sigma$ & $<1\sigma$ \\   
        $z = 120$ & $<1\sigma$ & $<1\sigma$ & $<1\sigma$ & $<1\sigma$ & $<1\sigma$ & $<1\sigma$ \\   
        $z = 130$ & $<1\sigma$ & $<1\sigma$ & $<1\sigma$ & $<1\sigma$ & $<1\sigma$ & $<1\sigma$ \\   
        $z = 150$ & $<1\sigma$ & $<1\sigma$ & $<1\sigma$ & $<1\sigma$ & $<1\sigma$ & $<1\sigma$ \\   
    \caption{The significance of the detection of 21\,cm power spectrum versus redshift for three concept arrays (the fiducial array from \citet{smith_and_pober_2025} and two hypothetical arrays with $10\times$ and $100\times$ higher sensitivity, respectively) and two foreground (FG) models (no foregrounds vs. the foreground wedge).  See \S\ref{sec:subtleties} for a discussion of the use of the total significance metric in this context.}
    \end{longtable}
    \label{tab:sensitivities}
\end{table*}

To gain some quantitative intuition about the effect of the wedge slope on the sensitivities of 21\,cm dark ages experiments, we use \texttt{21cmSense} v2 \citep{murray_et_al_2024} to determine the significance with which the 21\,cm power spectra can be detected for a range of redshifts.  A full description of the algorithms underlying \texttt{21cmSense} can be found in \citet{pober_et_al_2013a} and \citet{pober_et_al_2014}, while a concise pedagogical discussion of the approach can be found in \citet{liu_and_shaw_2020}.  Here we provide a brief summary of the key steps as relevant to the current work.  \texttt{21cmSense} estimates the full sampling pattern of an instrument, using both the antenna layout and basic observational parameters to construct a $(u,v)$ coverage map.  Combined with a choice of bandwidth and frequency resolution, the full 3D $k$-space sampling pattern is estimated (using Equations \ref{eq:kperp} and \ref{eq:kpar} to convert from observational parameters to cosmological $k$ space).  The 3D $k$ space sampling pattern is then used to determine a noise level based on the system temperature of the instrument, which is assumed to track the synchrotron spectrum of the sky \citep{thompson_et_al_2017} plus a small frequency-independent receiver contribution.  Lastly, this 3D sampling is binned into either 2D $(k_\perp, k_\parallel$) space or 1D $|k|$ space.  Furthermore, all the measurements can be combined into a total detection significance, averaging over any $k$-dependence in the signal-to-noise levels.  For simplicity, this is the number we present in this work; caveats associated with this metric are discussed in \S\ref{sec:subtleties}.

As a point of reference, we use the fiducial 21\,cm dark ages array design presented in \citet{smith_and_pober_2025}.  In brief, this array concept consists of 82,944 tightly packed dual-polarization 10\,m dipole antennas, for a total collecting area of $\sim 2.5\,\rm{km}^2$.  The dipoles are grouped together in $4\times4$ ``sub-arrays,'' which combine into a single element (much like the Murchison Widefield Array tile; \citealt{tingay_et_al_2013a}); this fiducial array thus has 5,184 correlated elements, easing the computational burden compared with a full correlation of every dipole.  \citet{smith_and_pober_2025} show that, in the absence of foregrounds, this fiducial array delivers a $10\sigma$ detection of the $z = 30$ 21\,cm power spectrum given a 5-year mission lifetime.

Table \ref{tab:sensitivities} presents an extension of this analysis, examining the significance of detection (i.e. the ``number of sigmas'') as a function of redshift.  (We discuss several subtleties regarding this metric in \S\ref{sec:subtleties}).  Note that all significances below $1\sigma$ are reported as ``$<1\sigma$'' to effectively convey a non-detection, but that the quantitative scaling trends discussed below versus redshift and foreground cut do continue to hold true.

We consider three different array concepts and two different foreground models.  For arrays, we use the fiducial array from \citet{smith_and_pober_2025} and two hypothetical arrays with $10\times$ and $100\times$ higher sensitivity, respectively.  We remain agnostic about how these sensitivity increases could be achieved.  \citet{smith_and_pober_2025} show that, as expected, the sensitivity scales linearly with observing time and roughly linearly with collecting area in the compact core, but closer dipole spacing and full correlation of elements (or at least smaller subarrays) could also boost sensitivity if engineering constraints can be overcome.  

For foreground models, we consider two simple cases: a no foreground case, and a model where all modes within the horizon line of wedge (defined by equation \ref{eq:horizon}) are excluded.  We note that neither of these are default modes of \texttt{21cmSense}, which uses an ``optimistic'' model that excludes only the region of $k$ space corresponding to foregrounds within the primary beam of the instrument (see \citet{pober_et_al_2016a} for a discussion of how location in the instrument's beam maps to 2D $k$ space), and a ``moderate'' model where foregrounds are assumed to extend past the horizon given by a constant additive ``buffer.''  As discussed in \S\ref{sec:background}, details about the exact distribution of foregrounds in (and beyond, in the case of the buffer) the wedge depend on analysis technique and array layout.  Furthermore, there is no definitive mapping between foreground subtraction techniques and specific modes of the 2D power spectrum that can be recovered.  Hence, we consider our two simple models as representative of best-case scenarios: the no-foreground model is simply the best one could ever hope to do, whereas the wedge-only model is the best one could hope to do with the foreground avoidance approaches used by many current ground-based 21\,cm EoR experiments.

\subsection{Subtleties of the Total Detection Significance Metric}
\label{sec:subtleties}

While the total detection significance (i.e. the number of sigmas) is a simple sensitivity metric, there are important caveats regarding its use in the present analysis.
First, although \citet{pober_et_al_2014} show that this value is largely correlated with scientific constraints at EoR redshifts, independent of the details of the array layout, that work considered independent frequency bands centered on each redshift bin; inside a bin, the power spectrum and noise are treated as constant.  In the Appendix of \citet{smith_and_pober_2025}, however, it is noted that an approximately co-eval bin (e.g. $\Delta z \lesssim0.5$) at dark ages redshifts spans a very narrow range of frequencies, and working with a very small bandwidth prevents the analysis of large-scale modes in $k_\parallel$.  As argued in \citet{smith_and_pober_2025}, there is significant cosmological information in these modes, but an analysis will need to forward model the mixing of information coming from multiple redshifts (which, in principle, can be done, especially since the dark ages 21\,cm signal is analytically tractable).  \texttt{21cmSense} does not support this forward modeling and we leave that problem to a future work; instead, we follow the same approach of \citet{smith_and_pober_2025} and use an 18\,MHz bandwidth centered on the redshift of interest for the sensitivity predictions.  In practice, this means that redshift bins in Table \ref{tab:sensitivities} have some overlap (especially at the highest redshifts, where the band also covers periods before the 21\,cm signal is in principle observable), which will need to be taken into account when considering the extraction of cosmological parameters.  However, neglecting this effect does not affect the key results of the present work about the impact of the wedge.

We also note that sample variance, which becomes the dominant source of uncertainty once thermal noise is appreciably reduced, is not included in these significances.  Sample variance will also need to be accounted for to determine the level of cosmological constraint possible with these experiments, but its inclusion is again not necessary to understand the impact of the wedge as we consider here.

\section{Discussion}
\label{sec:discussion}

There are two key takeaways from Table \ref{tab:sensitivities}.  First, it is clear that losing modes of the power spectrum within the foreground wedge consistently leads to a factor of $\sim10$ decrease in detection significance.\footnote{The reader may notice an apparent discrepancy with Figure 8 of \citet{smith_and_pober_2025}, which shows a $10\sigma$ dection for no foregrounds, but a $4.9\sigma$ and $0.41\sigma$ significance for ``foreground subtraction'' and ``foreground avoidance'', respectively.  These terms correspond to the optimistic and moderate foreground models with \texttt{21cmSense} and, as noted in \S\ref{sec:sensitivities}, both are distinct from our ``foreground wedge'' model (which is more akin to the moderate model, but lacks the additive buffer of spillover beyond the horizon).}  Comparison with Figure \ref{fig:wedge-slope} shows that this can be largely understood geometrically: at high redshifts, the wedge consistently occupies $\gtrsim 90\%$ of 2D $k$-space.  In terms of pure detection significance, this loss of modes can be compensated for with a more sensitive array, although some science cases are likely to intrinsically benefit from measuring larger numbers of modes.

Second, it is clear how steeply the significance of detection drops with increasing redshift.  This is not due simply to the increasing wedge slope, as it affects the no foreground models as well.  Rather, this effect is due to the increasing sky noise temperature, which we model as following the Galactic synchrotron radiation with a power law index of 2.55 vs. frequency.  Since the power spectrum scales as the square of the sky power, the noise sharply increases towards the highest redshifts.  Even in the no foreground case, our hypothetical array with 100 times the sensitivity of the \citet{smith_and_pober_2025} fiducial array cannot detect the signal above $z = 100$.  The effect of overlapping redshift bins discussed in \S\ref{sec:subtleties} may also make these already low sensitivities overly optimistic.

There are, however, two possible reasons for continued optimism.  First, dark ages cosmology experiments will need to operate in space to be above the ionosphere and, as noted in \citet{smith_and_pober_2025}, this is likely to aid foreground subtraction techniques (which, for ground-based experiments, must contend with ionospheric refraction and Faraday rotation causing time-variability in the received foreground emission).  Second, the sky temperatures used to calculate the noise levels at the highest redshifts may be overly large.  It has been known for some time that the Galactic plane becomes optically thick to low radio frequencies due to free-free absorption \citep{ellis_and_hamilton_1966,cane_and_whitham_1977,dwarakanath_2000,cong_et_al_2021}, meaning the overall sky power stops rising so steeply and may even start to become fainter.  Optically thick regions will also obscure the 21\,cm signal \citep{seitova_and_pober_2022}, so one must continue to make observations away from the Galactic plane; however, there may be a range of frequencies or fields where the Galactic plane contributes sizably to the system temperature through the sidelobes of the instrument primary beam and the overall reduction in system temperature outweighs the loss of lines of sight to the 21\,cm signal.  Further study is needed to determine just how this affects the detectability of the highest redshift 21\,cm signals.

\section{Conclusions}
\label{sec:conclusions}

We have presented an analysis of the impact of the foreground wedge on 21\,cm experiments targeting the cosmic dark ages.  Because the slope of the wedge grows steeper with increasing redshift, its impact is significant.  Using the foreground avoidance approach pioneered by ground-based EoR experiments results in approximately an order of magnitude loss of sensitivity, compared with a no-foreground case.  Excluding the foreground wedge from its measurements, even the 2.5\,km$^2$ fiducial array of \citet{smith_and_pober_2025} cannot detect the 21\,cm signal from $z=30$.  Since the intrinsic sensitivity of the measurements scales linearly with observing time and core collecting area, we would need to operate the fiducial array for 50 years or increase its footprint to $\sim25\,{\rm km}^2$ to gain back all that is lost to foregrounds.  While removing 100\% of foreground emission is likely impossible (some 21\,cm modes are almost certainly completely degenerate with the intrinsic foreground power), foreground subtraction is likely a requirement to enable dark ages 21\,cm cosmology with the scale of experiment that can be achieved in the next one to two decades.  Pushing to the higher redshifts in the heart of the cosmic dark ages likely also requires significant foreground mitigation compared with the approach of many current ground-based EoR experiments.

\begin{acknowledgments}
The authors acknowledge support for this work from NASA grants 80NSSC18K0389, 80NSSC21K0693, and 80NSSC22K1745.  WS also acknowledges support from a NASA Rhode Island Space Grant Fellowship. Part of this research was conducted using computational
resources and services at the Center for Computation and Visualization, Brown University. \end{acknowledgments}

%

\vspace{5mm}


\software{
21cmSense \citep{murray_et_al_2024}, 21cmFast \citep{mesinger_et_al_2011,murray_et_al_2020}, astropy \citep{astropy1,astropy2,astropy3}, lunarsky (\url{ https://github.com/aelanman/lunarsky}), pyuvdata \citep{hazelton_et_al_2017,keating_et_al_2025}
          }






\bibliography{poberbib}{}
\bibliographystyle{aasjournal}



\end{document}